\documentclass[english]{article}
\usepackage[T1]{fontenc}
\usepackage[latin9]{inputenc}
\usepackage{babel}

\begin{document}
THE FIRST J=1+ T=0 STATES in a SINGLE j SHELL CONFIGURATION in EVEV-EVEN
NUCLEI.

Larry Zamick

1.Department of Physics and Astronomy ,Rutgers University$^{1}$Pisctaway,New
Jersey 08854

Abstract

The first even-even nucleus for which there are J=1+ T=0 states in
a single j shell configuration is 48Cr. If we limit 

ourselves to single j there are no M1 transitons from these states
to any J=0+ T=0 states. 

.1. Absence of Spin J=1 states in j$^{4}$ Configurations

1a. J=1 T=2 States

1. Introduction

In early shell model calculations by McCullen et al.{[}1{]} and Ginocchio
and French {[}2{]}. it was noted that in the f$_{7/2}$shell 

certain combinations spin and isopin did not exist. For example there
were no J=0+T=1 states in $^{44}$Ti and no J=1+

states with T= T(min)+1 where T(min)= |N-Z|/2. There were also no
J=1+ states with T=T(max). However those states

are analogous to states of a system of idetical particles i.e. calcium
isotopes . Explanations for some of the missing 

states can be have been shown by low brow tecniques as wll be discussed
later.

In the mid eighties papers were published which counted the states
in a more systematic way. There include the works 

of I. Talmi on recusion relations for counting the states {[}3{]}
of identical fermionsand by Zhao and Arima {[}4{]} who 

obtained expressions for the number of T=0, 1 and 2 states for protons
and neutrons in a single j shell. Indeed the latter 

authors give all the answers to the counting questions addressed in
this paper.

.

2.. The first occurence of J=1+ T=0 states in the single j shell-48Cr

.

There are no J=1+ T=0 states for 4 nucleons in the single j shell.
To make things more concrete consider$^{^{44}}$Ti. The two f$_{7/2}$
protons can have angular momenta 0,2,4 and 6 all occuring once;likewise
the 2 neutrons. The (J$_{p}$,J$_{n}$) configurations that can add
up to a total J=1 are (2,2), (4,4) and (6,6).Thus there are three
J=1+ states.The possile isospins are 0, 1 and 2. Let us next consider
44Sc. The 3 neutrons can have angular momenta 3/2,5/2,7/2,9/2,11/2,
and 15/2 all occuring only once. The states that add up to one are
(7/2,5/2), (7/2,7/2) and (7/2,9/2).Again we have three states.However
since 44Sc has |T$_{z}$ =1 the isospins can only be one or two.Hence
there are no T=0 J=1+ states in 44Ti which are of the (f$_{7/2}$)$^{4}$
configuration.

We next consider 48Cr. The possible states of four protons , including
seniority labels are:

v=0 J=0

v=2 J= 2, 4, 6

v=4 J=2{*},4{*},5, 8

The possible J=1+ states are (2,2), (4,4), (6,6), (2{*},2{*}), (4{*},4{*}),
(5,5), (8,8), (2,2{*}), (2{*},2), (4,4{*}), (4{*},4), (4,5), (5,4),
(4{*},5), (5,4{*}), (5,6), (6,5). There are seventeen such states
with a priori possible isospins T=0,1,2,3,4. We next consider 48V
which consists of three protons and five neutrons.The latter can also
be regarded as three neutron holes so the possible states are the
same for neutrons and protons. The three proton states are 3/2,5/2,7/2,9/2,11/2,
and 15/2 all occuring only once. Te possible 1+ states are (3/2,3/2),
(5/2,5/2), (7/2,7/2), (9/2,9/2), (11/2,11/2), (15/2,15/2), (3/2,5/2),
(5/2,3/2), (5/2,7/2), (7/2,5/2), (7/2,9/2), (9/2,7/2), (9/2,11/2),
(11/2,9/2) . There are 14 such states and they all must have isospins
greater than zero. Hence the number of T=0 J=1+ states of the (f$_{7/2}$)$^{8}$
configuration is (17-14)= three.

The wave functions of these states is included in a larger compilation
by A.Escuderos, L.Zamick and B.F. Bayman {[}5{]}

.It i s there noted that becuase the both protons and neutrons are
at mid shell , the quantitiy s= (-1)$^{V}$ is a good quantum number
where V=( v$_{p}$+ v$_{n}$)/2. Referring to ref {[}1{]} for J=0+T=0
there are 4 staes with S=+1 and two with S=-1.All J=0+ T=1 states
have s=-1 while all T=2and T=4 states have s=+1.There are two J=1+
T=0 states wirh s=-1 at energies of 7.775 and 9.258 MeV.There is one
s=+1 state at 9.037 MeV with a rahter simple wave function 1/$\sqrt{2}$
{[}(4{*},5) + (5,4{*}){]}.The lowest J=0+ T=0 state has s= +1.

4. M1 Selection Rules 

There is a modern twist to what we are here doing. There has been
an extensive review of M1 excitations ,including spin flip modes ,scissors
modes e.t.c. by K.Heyde, P.Von Neumann-Cosel and A.Richter {[}6{]}.
The mode we are here considering ,has , to best of our knowledge,
not yet been studied experimentally. There have been studies of M1
T=0 to T=0 transitions e.g. the electro-excitation of T=0 J=1+ excited
states of 12C but these involve more than one shell {[}{]}. Isospin
impurites are very important for these transitions because the isovector
M1 coupling constants are much larger than the isoscalar ones. 

One simple selection rule for M1 transitions in this limited model
space is that M1(T=0$\rightarrow$T=0) equals zero.

.To see this we note that in the single j shell space we can replace
the M1 operator by g$_{_{j}}$J. The M1 matrix element for a T=0 to
T=0 transition is thus proportional to (g$_{j}$$_{\pi}$+ g$_{j}$$_{\nu}$)
, i.e. the isoscalar sum. But if such a term is non-zero it would
imply that the total angular momentum operator J (obtained by setting
the two g's above each equal to 1/2) could induce an M1 transition,which,of
course ,it cannot.

Another {}``midshell'' selection rule is that the quantum number''
s'' has to be the same for the initial J=1,T=0 state and for 

any final state e.g J=1+,T=1 or J=2+,T=1.

. Although not necessary it is nevertheless instructive to show in
more detail why the T=0 to T=0 matrix element vanishes. Consider a
transition from s=-1 to s=-1. In the wave functions there will be
no amplitude of the configuration

(Jp,Jn) =(2,2) but there will be of (2,2{*}) and (2{*},2) The transition
matrix element will have the form

<(2,2{*})$^{2}$ +(2{*},2)$^{2}$|| M1 || (2,2{*})$^{1}$- (2{*},2)$^{1}$>.
This is equal to<(2,2{*})$^{2}$||M1|| (2,2{*})$^{1}$> - <(2{*},2)$^{2}$||M1||
(2{*},2)$^{1}$>

.Since,in the single j shell one can replace M1 by g$_{j}$ J the
matrix element <2||M1||2> is equal to <2{*}|| M1 ||2{*}>

.We thus see that the complete  matrix element vanishes.

.

.{[}1{]} J.D. Mcullen,B.F. Bayman and L. Zamick ,Phys. Rev. 134 ,B515
1964)

{[}2{]} J.N. Ginocchio and J.B. French Phys.Lett. 7,137 (1963)

{[}3{]} I.Talmi Phys. Rev. C72, 037302 (2005)

{[}4{]}.Y.M.Zhao and A. Arima Phys. Rev. C71,047304 (2005);Phys. Rev.
C72 ,064333 (2005)

{[}5{]} A. Escuderos,L.Zamick and B.F. Bayman,LANL manuscript arXivnucl-th/0506050
(2007)

{[}6{]} K. Heyde, P. Von Neumann-Cosel and A. Richter Rev. Mod. Phys.
82,2365 (2010) 
\end{document}